\documentclass[prl,showpacs,twocolumn]{revtex4}
\usepackage{graphicx}
\usepackage{psfig}
\begin{document}
\def\/{\over}
\def\<{\langle}
\def\>{\rangle}
\def\({\left(}
\def\){\right)}
\def\[{\left[}
\def\]{\right]}
\def\I{{\cal I}}
\def\i{{\rm i}}
\def\d{{\rm d}}
\def\e{{\rm e}}
\title{Scars on quantum networks ignore the Lyapunov exponent}
\author{Holger Schanz}
\author{Tsampikos Kottos}
\affiliation{
Institut f{\"u}r Nichtlineare Dynamik\mbox{,} Universit{\"a}t G{\"o}ttingen 
and Max-Planck-Institut f\"ur Str\"omungsforschung,
Bunsenstra{\ss}e 10, D-37073 G\"ottingen, Germany
}
\date{\today}
\pacs{03.65.Sq, 05.45.Mt}
\begin{abstract}
  We show that enhanced wavefunction localization due to the presence of short
  unstable orbits and strong scarring can rely on completely different
  mechanisms. Specifically we find that in quantum networks the shortest and
  most stable orbits do not support visible scars, although they are
  responsible for enhanced localization in the majority of the
  eigenstates. Scarring orbits are selected by a criterion which does not
  involve the classical Lyapunov exponent. We obtain predictions for the
  energies of visible scars and the distributions of scarring
  strengths and inverse participation ratios.
\end{abstract}
\maketitle
One of the most striking ways in which the underlying classical dynamics of a
chaotic system manifests itself in the corresponding quantum behavior is the
{\it scar} phenomenon
\cite{Hel84,KH98,Bog88,AF94,Kap01,Sri91,W+96,BAF01,KAG01,G+02}.  A scar is a
quantum eigenfunction with excess density near an unstable classical periodic
orbit (PO). Such states are not expected within Random-Matrix Theory (RMT),
which predicts that wavefunctions must be evenly distributed over phase space,
up to quantum fluctuations \cite{Mehta}. Experimental evidence and
applications of scars come from systems as diverse as microwave resonators
\cite{Sri91}, quantum wells in a magnetic field \cite{W+96}, Faraday waves in
confined geometries \cite{KAG01}, open quantum dots \cite{BAF01} and
semiconductor diode lasers \cite{G+02}.

\begin{figure}[!t]
\includegraphics[scale=0.5]{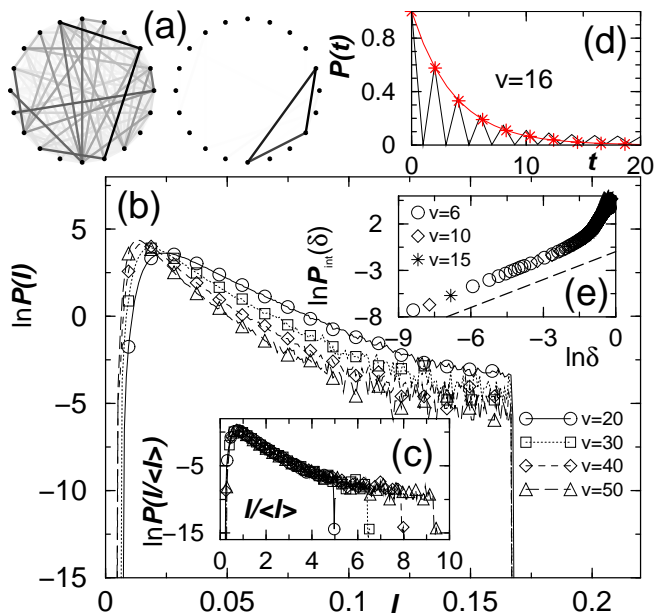}
\caption{\label{figIPR}
  Statistical properties of eigenstates for fully connected graphs with
  different valencies $v=V-1$. The eigenstates were obtained by
  diagonalization of the bond-scattering matrix Eq.~(\protect\ref{bsm}), and a
  statistical ensemble was generated by choosing random bond lengths. (a) The
  bond intensities of two representative eigenstates are shown with a
  gray-scale for a typical state (left) and a scar on a triangular PO (right).
  (b) Probability distribution of the IPR, showing a step-like cutoff at
  $\I=1/6$ that can be attributed to scarring on triangular orbits. (c) The
  bulk of the IPR distribution shows scaling according to Eq.~(\ref{bulkI}).
  (d) The quantum return probability and a classical approximation ($\ast $)
  based only on the period-two orbits; (e) Rescaled and integrated
  distribution of triangle scars ${\cal P}_{\rm int}
  (\delta)=B^{-1}\,\int_0^\delta\d\delta ' {\cal P}^{(3)}(\delta ')$. The
  dashed line has slope $1$, corresponding to the theoretical prediction
  Eq.~(\ref{pdelta}).}
\end{figure}

Quantum networks (graphs) are established models in the field of mesoscopic
physics, from which most of the above examples are drawn, as well as in many
other areas including molecular and mathematical physics and quantum
computation (see \cite{Kuc02,ACMProc,KS97,KS99} and Refs.\ therein).  In
recent years they have become one of the most prominent tools in quantum chaos
because they allow to study with simple means the applicability of RMT and its
limitations due to system-specific properties \cite{KS97,KS99,BBK01}.  For
example, it was shown recently \cite{Kap01} that statistics of the bulk of
graph eigenfunctions (including, e.~g.\, the left eigenstate in
Fig.~\ref{figIPR}a) conform with the existing theories describing the effect
of short unstable periodic orbits on the localization properties of
wavefunctions.

Therefore it is surprising that the same does not apply to the small but
important group of strongly scarred eigenstates (Fig.~\ref{figIPR}a, right),
which we study in this letter. We show that, contrary to common intuition and
accepted theories \cite{Hel84,KH98,Bog88,AF94,Kap01}, the shortest and least
unstable orbits of the system produce almost no visible scars, although they
are responsible for enhanced localization within the bulk of states. We derive
a condition, Eq.~(\ref{criterion}) below, selecting orbits relevant for strong
scarring from topological information only and without any reference to the
classical Lyapunov exponent.  Based on this insight we are able to give a
criterion for the energies at which strong scars are to be expected,
Eq.~(\ref{quantization}), and describe their statistical distribution,
Eq.~(\ref{pdelta}). In view of the numerous and diverse applications of
quantum networks \cite{Kuc02,ACMProc,KS97,KS99,BBK01}, these main results
should be of broad interest in their own right.  On top of this, some
important conclusions generalize beyond graphs. In particular, our results
provide clear evidence for the fact that {\em enhanced wavefunction
  localization due to the presence of short unstable orbits and strong
  scarring can in principle rely on completely unrelated mechanisms} and can
also leave distinct traces in statistical measures such as the distribution of
inverse participation ratios (IPR).

Following the quantization outlined in \cite{KS97} we consider graphs which
consist of $V$ vertices connected by a set of $B$ bonds.  The number of bonds
emanating from a vertex $j$ is its valency $v_j$.  A basis state on the network
is specified by a {\em directed} bond $d=[i\to j]$, i.~e.\ an ordered pair of
connected vertices $i$, $j$. Hence a quantum wavefunction is just a set of
$2B$ complex amplitudes $a_{d}$, normalized according to
$\sum_{d}|a_{d}|^2=1$. The standard localization measure is the IPR
\begin{equation}
\I=\sum_{d=1}^{2B} |a_d|^4\,.  
\end{equation}
Ergodic states which occupy each directed
bond with the same probability have ${\cal I}= 1/2B$ and up to a constant
factor depending on the presence of symmetries this is also the RMT
prediction. In the other extreme $\I =1$ indicates a state which is restricted
to a single bond only, i.~e.\ the greatest possible degree of localization.

Fig.~\ref{figIPR}b shows the distribution of $\I$ for fully connected graphs.
Some features of this distribution are explained by the original scar theory
of Heller \cite{Hel84} and extensions of it \cite{KH98,Kap01}. The main idea
is to connect localization properties of eigenfunctions to the dynamics of the
system. For example, the identity $\<\I\>=\lim_{T \rightarrow \infty} {1\over
  T}\int_0^T\d t\, \<P(t)\>$ expresses the mean IPR in terms of the quantum
return probability (RP) $P(t)$, averaged over time and initial states. It is
then argued that the short-time dynamics, approximated semiclassically with
some short PO's, provides sufficient information for estimating $\<\I\>$.
Within this approach it is clear that the orbits with the lowest Lyapunov
exponent (LE) have the largest influence on eigenfunction localization because
classical trajectories can cycle in their vicinity for a relatively long time
and increase the RP beyond the ergodic average. For the graphs studied in this
letter, some PO's $p$ and their LE $\Lambda_{p}$ are listed in
Table~\ref{table:po}. The shortest PO's have period $2$ and bounce back and
forth between two vertices. For large graphs $v\to\infty$ these are by far the
least unstable ones, as their LE approaches 0 while all others become
increasingly unstable $\Lambda_{p}\sim\ln v$. Indeed the period $2$ orbits
totally dominate the classical and quantum RP at short times (see
Fig.~\ref{figIPR}d). Including the contribution of these orbits only, Kaplan
obtained a mean IPR which is by a factor $\sim v$ larger than the RMT
expectation, in agreement with numerics \cite{Kap01}. Moreover, following the
same line of argumentation as in \cite{KH98} we get that the bulk of the IPR
distribution scales as 
\begin{equation}
\label{bulkI}
\tilde P(\I/\<\I\>)=\<\I\>\,P(\I)
\end{equation}
indicating that 
the whole bulk of ${\cal P}(\I)$ is
effectively determined by the least unstable orbits (Fig.~\ref{figIPR}c).

\begin{table}[!t]
\caption{
The topology of the shortest PO's of a fully connected graph with valency 
$v$ are shown with the corresponding amplitude, Lyapunov exponent, and IPR.
\label{table:po}}
\unitlength0.3mm
\linethickness{\unitlength}
\begin{ruledtabular}
\begin{tabular*}{\hsize}{
l@{\extracolsep{0ptplus1fil}}
c@{\extracolsep{0ptplus1fil}}
c@{\extracolsep{0ptplus1fil}}
c@{\extracolsep{0ptplus1fil}}
c@{\extracolsep{0ptplus1fil}}
c@{\extracolsep{0ptplus1fil}}
c}
$t_{p}$&$p$&$A_{p}$&$\Lambda_{p}(v\to\infty)$&$\I_{p}$ \\
\colrule
2&
\begin{picture}(20,20)(-10,-10)
\put(-8,-6){\circle*{5}}
\put(-8,-6){\line(1,0){16}}
\put(+8,-6){\circle*{5}}
\end{picture}
&$(2/v-1)^{2}$
&$4/v$
&1/2
\\
3&
\begin{picture}(20,20)(-10,-10)
\linethickness{10\unitlength}
\put(-8,-8){\line(1,+2){8}}
\put( 0,+8){\line(1,-2){8}}
\linethickness{\unitlength}
\put(-8,-8){\line(1,0){16}}
\put(-8,-8){\circle*{5}}
\put( 0,+8){\circle*{5}}
\put(+8,-8){\circle*{5}}
\end{picture}
&$(2/v)^3$ 
&$2\ln v$
&1/6
\\
4&
\begin{picture}(20,20)(-10,-10)
\put(-8,+8){\line(1,-2){8}}
\put( 0,-8){\line(1,+2){8}}
\put(-8,+8){\circle*{5}}
\put( 0,-8){\circle*{5}}
\put(+8,+8){\circle*{5}}
\end{picture}
&$(2/v)^2(2/v-1)^{2}$
&$\ln v$
&1/4
\\
&\begin{picture}(20,20)(-10,-10)
\put(-8,-8){\line(1,0){16}}
\put(+8,-8){\line(0,1){16}}
\put(+8,+8){\line(-1,0){16}}
\put(-8,+8){\line(0,-1){16}}
\put(-8,-8){\circle*{5}}
\put(+8,-8){\circle*{5}}
\put(+8,+8){\circle*{5}}
\put(-8,+8){\circle*{5}}
\end{picture}
&$(2/v)^{4}$ 
&$2\ln v$
&1/8
\end{tabular*}
\end{ruledtabular}
\end{table}

With all this evidence for their prominent role in wavefunction localization,
one clearly expects to see strong scarring on the period 2 orbits
\footnote{Apparently this expectation is also supported by other scar theories
  which up to now have not been applied to graphs. For example, Agam and
  Fishman \protect\cite{AF94} predicted the positions of visible scars by a PO
  sum in which the orbits are weighed according to their stability.}. Such
states would essentially be concentrated on two directed bonds and give rise
to $\I\sim 1/2$. However, in this region ${\cal P}(\I)$ is negligible. {\em
  The shortest and least unstable orbits of our system produce no visible
  scars.}  Note that the same applies also to the value $\I=1/4$ expected from
the V-shaped orbits of Table~\ref{table:po}.  In fact ${\cal P}(\I)$ has an
appreciable value only for $\I\le 1/6$ (Fig.~\ref{figIPR}b). The position of
this cutoff precisely coincides with the IPR expected for states which are
scarred by triangular orbits.  They occupy six directed bonds since, due to
time-reversal symmetry, scarring on a PO and its reversed must coincide.
Indeed a closer inspection shows that the vast majority of states at
$\I\approx 1/6$ look like the example shown in Fig.~\ref{figIPR}a (right).  Of
course the step at $\I=1/6$, which is present for any graph size $V$, is
incompatible with the scaling of $P(\I)$ mentioned above and indeed this
relation breaks down in the tails at the expected points (inset of
Fig.~\ref{figIPR}c). We conclude that {\em visible scars on short unstable
  orbits can strongly modify the tails of the IPR distribution} even beyond
the known predictions for the influence of short PO's on wavefunction
localization.

In the rest of the paper we will formulate a theory which explains the above
observations. To this end we must be more explicit concerning the dynamics on
the graph.  We consider particles with fixed wavenumber $k$, propagating on
the bonds and scattering at the vertices. During the free propagation on a
directed bond the wavefunction accumulates a phase $kL_{ij}$, where
$L_{ij}=L_{ji}$ denotes the length of the corresponding bond.  At the vertices
current conservation and continuity of the wave function are required. These
boundary conditions can be translated into vertex scattering matrices, which
describe a unitary transformation from $v_{i}$ incoming to $v_{i}$ outgoing
waves at each vertex.  Without loss of generality we restrict the presentation
to the simplest case of {\em Neumann} boundary conditions \cite{KS97}, where
the scattering matrix of vertex $i$ is
\begin{equation}
\label{smatrix}
\sigma^{(i)}_{j,j'} = {2/ v_i}-\delta _{jj'}\,. 
\end{equation}
We can now combine the free propagation and the vertex scattering into a
$2B\times 2B$ operator, the bond-scattering matrix $S$ \cite{KS97}, which acts
on the amplitudes $a_{d}$ associated with the directed bonds. The matrix
element
\begin{equation}
\label{bsm}
S_{m\to n,i\to j}=\delta_{mj}\,(2/v_{j}-\delta_{in})\,\e^{\i kL_{ij}}
\end{equation}
describes a transition from the directed bond $d=[i\to j]$ to $d'=[m\to n]$.
We interpret $S$ as quantum time-evolution operator on the graph.
$(S^{t})_{d'd}$ is the complex probability amplitude to be after $t=0,1,\dots$
time steps on the directed bond $d'$ if the initial state was on $d$. In
particular, $|(S^{t})_{dd}|^{2}$ is the quantum RP shown in
Fig.~\ref{figIPR}b. $(S^{t})_{dd}=\sum_{p}A_{p}\e^{\i kL_{p}}$ is expanded as
a sum over all PO's of period $t$ starting at $d$.  Here $L_{p}$ is the total
length of orbit $p$. Assuming for simplicity $v_{j}\equiv v$ throughout the
graph, we express the amplitude $A_{p}$ by the number $r_{p}$ of reflections
along $p$, $A_{p}= (2/v)^{t-r_{p}}(2/v-1)^{r_{p}}$ (cf Table~\ref{table:po}).
The classical RP is obtained by summing, instead of the amplitudes, the
probabilities $M_{p}=|A_{p}|^2$. As $M_{p}<1$, the probability to follow the
PO decreases exponentially with time and the orbit is unstable.  Hence one
defines the {\em Lyapunov exponent} of a PO $p$ on a graph by
\begin{equation}
\label{LE}
\Lambda_{p}=-t_{p}^{-1}\ln M_{p}\,.
\end{equation}

Let us now come back to the problem of scarring and investigate the conditions
under which we can construct {\em perfect scars} on the graph, i.~e.
eigenstates $S|a\>= \e^{\i\lambda}|a\>$ which have the property that they have
non-zero amplitude only on a PO $p$ and vanish on all other bonds. Consider an
arbitrary vertex $j$ on $p$ and let $D^{(\pm)}_{j,p}$ be the set of directed
bonds which are leaving/approaching $j$ and belong to $p$. Similarly let
$\widehat D^{(\pm)}_{j,p}$ be the set of bonds {\em not} belonging to $p$. By
construction, the amplitude of $|a\>$ on these latter bonds vanishes, i.~e.\ 
all waves arriving on the bonds $D^{(-)}_{j,p}$ and transmitted across the
vertex to a bond in $\widehat D^{(+)}_{j,p}$ cancel each other
\begin{equation}
\label{zero}
0=\sum_{d'}S_{dd'}a_{d'}=
{2\over v_j}
\sum_{d'\in D^{(-)}_{j,p}} \e^{\i kL_{d'}}\,a_{d'}
\quad(d\in \widehat D^{(+)}_{j,p})\,.
\end{equation}
This equation has an important consequence: a perfect scar cannot live on a
single bond attached to $j$ because there would be no way to cancel the
transmitted wave. An exception are only vertices with valency $v_j=1$, for
which $\widehat D^{(\pm)}_{j,p}$ is empty. Formally a necessary {\em
condition for scarring orbits} $p$ is
\begin{equation}\label{criterion}
v_{j,p}\ge 2-\delta_{v_{j},1}
\qquad (\forall j\in p)
\end{equation}
where $v_{j,p}$ denotes the number of bonds attached to $j$ and belonging to
$p$.  Eq.~(\ref{criterion}) excludes, in particular, perfect scars on the
period-two orbits.  Applying the same reasoning that lead to Eq.~(\ref{zero})
now to the bonds $D^{(+)}_{j,p}$ and making use of Eq.~(\ref{smatrix}) we get
\begin{eqnarray}
\label{condition}
\e^{\i\lambda}\,a_{d}
=-\e^{\i kL_{d}}\,a_{\hat d} \qquad(d\in D^{(+)}_{j,p})\,,
\end{eqnarray}
which relates the amplitude on a directed bond $d\in p$ to the amplitude on
the reversed bond $\hat d\in \hat p$. This means that the same states are
scarred on $p$ and $\hat p$, as expected from time-reversal symmetry.
Substituting $d\Rightarrow \hat d$ we get $\e^{\i \lambda}\,a_{\hat d}=-\e^{\i
  kL_{d}}\,a_{d}$ which together with Eq.~(\ref{condition}) implies
\begin{equation}
\label{quantization}
(kL_{d}-\lambda)\,\mbox{mod}\,\pi=0\qquad\forall\, d\in p
\end{equation}
with arbitrary $\lambda$ \footnote{For $\lambda=0$ the scarred states satisfy
  the secular equation for graphs \protect\cite{KS97}. We have checked that
  our results remain valid in this special case.}.  Eq.~(\ref{quantization})
is a necessary and sufficient condition for the energies of perfect scars. It
is reminiscent of a simple Bohr-Sommerfeld quantization condition
$kL_{p}=2n\pi+\lambda$, as it applies, e.~g., to strong scars in billiards.
However, there is an important difference: not only does
Eq.~(\ref{quantization}) require quantization of the total action $kL_{p}$ of
the scarred orbit, it also implies action quantization on all the visited
bonds $d$. This stronger condition can only be met if the lengths of all bonds
on $p$ are rationally related. As in general the bond lengths are
incommensurate {\em there are no perfect scars for generic graphs}.

Nevertheless, for incommensurate bond lengths Eq.~(\ref{quantization}) can be
approximated with any given precision and then visible scars are expected. A
natural measure for the quality of a scarred state $|a\>$ is the total
probability $\delta_p=\sum_{d\notin p}|a_d|^2$ to find this state away from
the scarring orbit ($\delta_{p}=0$ corresponds to a perfect scar). We will
derive the probability density of strong visible scars ${\cal P}(\delta_{p}\to
0)$. We
represent the bond-scattering matrix $S$ as perturbation of a matrix $S_{0}$
which has a perfect scar on $p$, i.e.
\begin{equation}
\label{perturb}
S=\e^{+\i\varepsilon\Phi}\,S_{0} \approx (1+\i\varepsilon \Phi) S_{0}\,.
\end{equation}
The deviations from exact quantization for the individual bonds $d\in p$ have
been combined into a diagonal matrix $\Phi$ with
$\varepsilon\Phi_{dd}=(kL_{d}-\lambda)\,\mbox{mod}\,\pi$ for $d\in p,\hat p$
and $\Phi_{dd}=0$ otherwise. The strength of the perturbation is given by
$\varepsilon_{p}=\min_\lambda\max_{d\in
p}|(kL_{d}-\lambda)\,\mbox{mod}\,\pi|$.  For a PO covering $N$ undirected
bonds of the graph $N$ bond phases $kL_{d}\,\mbox{mod}\,\pi$ must
approximately coincide. Upon variation of $k$, they are independent and
uniformly distributed random numbers in $[-\pi,+\pi]$. Therefore the
probability density of a small perturbation is $p(\varepsilon\to
0)\sim\varepsilon^{N-2}$. From first order perturbation theory we have
\begin{equation}
\label{wfopt}
\delta = \varepsilon^2 
\sum_{d\notin p}\left |\sum_{m\ne n}
{\<a_m^{(0)}|\hat\Phi |a_n^{(0)}\>
\<d |a_m^{(0)}\>
\over 1-\e^{\i(\lambda_m^{(0)}-\lambda_n^{(0)})}}\right|^2\,,
\end{equation}
where $\lambda_m^{(0)}$, $|a_m^{(0)}\>$ are eigenphases and eigenvectors of
$S_0$, including the perfect scar $|a_n^{(0)}\>$.  The quantity
$x=\delta/\varepsilon^2$ is distributed with some probability density $\tilde
p(x)$ that is independent on $\varepsilon$. Consequently we have
$p(\delta|\varepsilon)= \varepsilon^{-2}\,\tilde p(\delta/\varepsilon^2)$. We
can now use ${\cal P}(\delta)=
\int\d\varepsilon\,p(\delta|\varepsilon)\,p(\varepsilon)$ together with
$p(\varepsilon\to 0)\sim\varepsilon^{N-2}$ to deduce
\begin{equation}
\label{pdelta}
{\cal P}^{(N)}(\delta)=C\,\delta^{(N-3)/2}\qquad(\delta\to 0)\,.
\end{equation}
Here $C=\int\d x\,x^{-(N-1)/2}\,\tilde p(x)$ is a constant which depends on
the size and topology of the graph. Note that, according to Eq.~(\ref{wfopt}),
$x$ is a sum of $B-N$ independent non-negative terms $x=\sum_{d\notin
  p}|x_d|^2$. Hence, $\tilde p(x)$ vanishes as $x^{B-N}$ for $x\to 0$ and as a
consequence the above integral $C$ exists. 

For fully connected graphs, triangles are the shortest orbits compatible with
Eq.~(\ref{criterion}) and we have ${\cal P}^{(3)}(\delta\to 0)=C$, i.~e.\ the
probability of scarring does not depend on the required quality of the scar.
This is in excellent agreement with numerical results (Fig.~\ref{figIPR}e) and
compatible with the step-like cutoff in ${\cal P}(\I)$ \footnote{Strictly
  speaking, $\delta$ gives only a lower bound for $\I$, $\I\ge 1/6-\delta/3$,
  such that the step at $\I=1/6$ is slightly smoothed. This is beyond the
  resolution of Fig.~\ref{figIPR}b).}.  In a similar way we can describe the
statistics of scars on other orbits. For example, according to
Eq.~(\ref{criterion}), the square-shaped orbits of Table~\ref{table:po} can support
scars and we did observe such states. However, Eq.~(\ref{pdelta}) gives ${\cal
  P}^{(4)}(\delta\to 0)\sim\delta^{1/2}$, i.~e.\ the probability of strong
square scars is much smaller than for triangles and consequently they leave no
distinct trace in ${\cal P}(\I)$.

\begin{figure}[!t]
\includegraphics[scale=0.5]{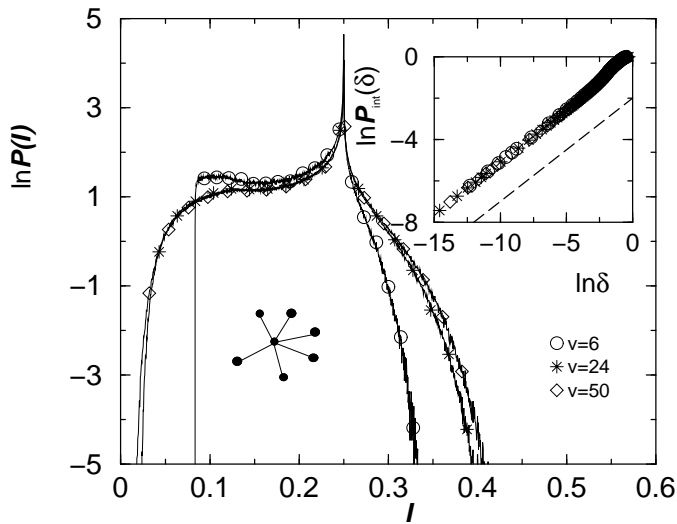}
\caption{\label{figstar}
  ${\cal P}(\I)$ for star-graphs of various valencies (inset a star-graph with
  $v=6$).  In the upper inset we report the integrated distribution ${\cal
    P}_{\rm int}(\delta)= \int_0^\delta d\delta 'P(\delta ')$ in a
  double-logarithmic plot. The dashed line has slope $0.5$ which corresponds
  to the theoretical prediction Eq.~(\ref{pdelta}).}
\end{figure}

In contrast to fully connected graphs, Eq.~(\ref{criterion}) allows in star
graphs scarring on the $V$-shaped orbits of Table~\ref{table:po}, because the
outer vertices have valency 1.  Applying Eq.~(\ref{pdelta}) in this case we
get ${\cal P}(\delta\to 0)\sim\delta^{-1/2}$, i.~e.  scarring is strongly
enhanced. As a consequence ${\cal P}(\I)$ is here totally dominated by scars,
showing a strong maximum at $\I=0.25$ (Fig.~\ref{figstar}). It would be very
interesting to relate this fact to spectral statistics, which for star graphs
corresponds to pseudo-integrable instead of chaotic classical dynamics
\cite{BBK01}.

\acknowledgments Discussions with E.~J.~Heller, S.~Fishman and
U.~Smilansky are gratefully acknowledged. T.~K.~acknowledges support by a
Grant from the GIF, the German-Israeli Foundation for Scientific Research and
Development.

\end{document}